\documentclass[fleqn,usenatbib,useAMS]{mnras}

\usepackage{mathptmx}

\usepackage[T1]{fontenc}
\usepackage{ae,aecompl}
\usepackage{blindtext}
\usepackage{adjustbox}
\usepackage{subcaption}
\usepackage[section]{placeins}
\DeclareGraphicsExtensions{.pdf,.png,.jpg,.eps}
\usepackage{epstopdf}
\usepackage{xcolor}
\usepackage[utf8]{inputenc}
\usepackage[T1]{fontenc}
\usepackage{lmodern} 
\usepackage{diagbox}

\usepackage{amsmath}
\usepackage{amssymb}
\usepackage{amsbsy}
\usepackage{comment}
\usepackage[normalem]{ulem}
\usepackage{enumitem}

\makeatletter
\usepackage{enumitem}
\newlist{mylist}{itemize}{4}
\setlist[mylist]{label=\textbullet,leftmargin=1.8em,%
  nolistsep,labelsep=1em,labelwidth=*,%
  topsep=-\parskip, after*={\@topsepadd\parskip}}
\makeatother

\makeatletter
\newcommand*{\rom}[1]{\expandafter\romannumeral #1}
\makeatother

\newcommand{\green}[1]{\iffalse #1 \fi}
\usepackage{pbox}

\definecolor{blue}{rgb}{0,0,1}

\usepackage{graphicx}	
\usepackage{float}
\usepackage{amsmath}	
\usepackage{amssymb}	
\usepackage{bm}		
\usepackage{pdflscape}	
\usepackage{xcolor}
\usepackage{exscale}
\usepackage{natbib}
\usepackage{rotating}
\usepackage{rotate}
\usepackage{afterpage}
\usepackage{booktabs,threeparttable}
\usepackage{relsize}
\usepackage{lscape}
\usepackage{tabularx}
\usepackage{longtable}
\usepackage{ltxtable}
\usepackage{newtxtext}
\usepackage{bbm}
\usepackage{afterpage}
\usepackage{bbold}
\usepackage{footnote}

\title{Mitigating baryonic effects with a theoretical error covariance}

\newcommand{\eg}{e.g$.$ }

\newcommand{\ie}{i.e$.$ }

\author[Maria G. Moreira et al.]{Maria G. Moreira,$^{1,2}$
Felipe Andrade-Oliveira, $^{1,2}$
Xiao Fang, $^{3}$
Hung-Jin Huang, $^{3}$
\newauthor Elisabeth Krause,$^{3,4}$
Vivian Miranda, $^{3}$
Rogerio Rosenfeld, $^{1,2,5}$
Marko Simonovi\'{c} $^{6}$
\\
$^{1}$Instituto de Física Teórica, Universidade Estadual
Paulista, São Paulo, Brazil \\
$^{2}$ Laboratório Interinstitucional de e-Astronomia - LIneA,
Rua Gal. José Cristino 77, Rio de Janeiro, RJ - 20921-400,
Brazil\\
$^{3}$ Department of Astronomy/Steward Observatory, University of Arizona, 933 North Cherry Avenue, Tucson, AZ
85721, USA \\
$^4$ Department of Physics, University of Arizona, 1118 E. Fourth Street, Tucson, AZ, 85721, USA\\
$^{5}$ 
ICTP South American Institute for Fundamental Research,
São Paulo, Brazil \\
$^{6}$ Theoretical Physics Department, CERN,
1 Esplanade des Particules, Geneva 23, CH-1211, Switzerland
}

\begin{document}
\maketitle

\begin{abstract}
 One of the primary sources of uncertainties in
 modeling the cosmic-shear power spectrum on small scales
 is the effect of baryonic physics.
 Accurate cosmology for Stage-IV surveys requires knowledge of the matter power spectrum deep in the nonlinear regime at the percent level. Therefore, it is important to develop reliable mitigation techniques to take into account baryonic uncertainties if information from small scales is to be considered in the cosmological analysis.
 In this work, we develop a new mitigation method for dealing with baryonic physics for the case of the shear angular power spectrum.
 The method is based on an extended covariance matrix that incorporates baryonic uncertainties informed by hydrodynamical simulations.
 We use the results from 13 hydrodynamical simulations and the residual errors arising from a fit to a $\Lambda$CDM model using the extended halo model code {\tt HMCode}  to account for baryonic physics. These residual errors are used to model a so-called theoretical error covariance matrix that is added to the original covariance matrix.
 In order to assess the performance of the method, we use the 2D tomographic shear from four hydrodynamical simulations that have different extremes of baryonic parameters as mock data and run a likelihood analysis comparing the residual bias on $\Omega_m$ and $\sigma_8$ of our method and the {\tt HMCode} for an LSST-like survey.
 We use different modelling of the theoretical error covariance matrix to test the robustness of the method. We show that it is possible to reduce the bias in the determination of the tested cosmological parameters at the price of a modest decrease in the precision. 
\end{abstract}


\section{Introduction}
One of the goals of modern cosmology is to uncover the nature of dark matter and dark energy.  
Current and new instruments aim at obtaining data with increasing quality and quantity. 
Surveys of galaxies such as the Extended
Baryon Oscillation Spectroscopic Survey (eBOSS\footnote{\tt
www.sdss.org/surveys/eboss}, \citealt{eBOSS20}) and the previous phases of the Sloan Digital
Sky Survey (SDSS, \citealt{SDSS-IV, SDSS-III}), the Hyper Suprime-Cam Subaru Strategic Program
(HSC-SSP\footnote{\tt hsc.mtk.nao.ac.jp/ssp}, \citealt{Hikage19}), the Kilo-Degree Survey
(KiDS\footnote{\tt kids.strw.leidenuniv.nl}, \citealt{Heymans21, Hildebrandt17}) and the Dark Energy Survey
(DES\footnote{\tt www.darkenergysurvey.org}, \citealt{DESY1_3x2pt}) 
have already delivered an outstanding amount of results.
And future surveys, such as the
Dark Energy Spectroscopic Instrument (DESI\footnote{\tt
www.desi.lbl.gov}, \citealt{DESI16}), the Vera Rubin Observatory Legacy Survey of Space and
Time (LSST\footnote{\tt www.lsst.org}, \citealt{LSST19}), Euclid\footnote{\tt
www.euclid-ec.org} \citep{Euclid11} and the Nancy Grace Roman Space Telescope\footnote{\tt roman.gsfc.nasa.gov} \citep{Spergel15} will provide even more accurate information.

In order to extract cosmological information from these data it is important to have an accurate theoretical modelling of the measured observables. One of the key obstacles in the interpretation of weak lensing measurements is the modelling of baryonic feedback at small scales. For a recent review of the challenges of baryonic feedback and relevant references see, \eg \citet{Chisari:2019tus}.

State-of-the-art hydrodynamical simulations allow the study of the impact of baryonic feedback galaxy-formation dynamics on the matter power spectrum. 
However, these simulations can not predict the behaviour of feedback processes from first principles and several phenomenological parameters must be assumed. 
Therefore, there are uncertainties in the predictions of baryonic feedback from
these simulations since there is a range of different values for these parameters that can be used.

The main methods used for mitigating baryonic uncertainties consist on using the information from current cosmological hydrodynamical simulations to:\\
$\bullet$ choose a scale ("scale-cut") below which one can not trust the theoretical modelling of the power spectrum

\citep{Krause2017}; \\ 
$\bullet$ use principal component analysis to find the modes that are the most significant in describing baryonic impacts and marginalize over them \citep{Eifler:2014iva, Huang21}; \\
$\bullet$ construct self-calibrated phenomenological models that mimic the baryonic effects in the structure of dark matter halos \citep{Mead:2015yca}. 

DES and HSC have chosen to mitigate baryonic uncertainties using scale cuts to eliminate the impact of baryonic physics as modelled by either
the OWLS hydrodynamical simulations \citep{vanDaalen11} in the case of DES \citep{Krause2017}, or by a modification of the dark matter power spectrum due to the
AGN feedback modelled by a fitting function in the case of HSC \citep{Hamana:2019etx,Hikage19}. KiDS, on the other hand, uses in their fiducial analysis a baryon feedback parameter \citep{Asgari_2021}.

In a recent study \citet{Huang:2018wpy} showed that in the case of weak lensing for the LSST there are still residual errors using these mitigation methods.
In this paper, we aim to mitigate these residual errors, particularly the ones arising from a likelihood analysis using hydrodynamical simulations as input data and a theoretical model using {\tt HMCode} \citep{Mead:2015yca}.

In order to develop a new mitigation method for the residual errors in
an LSST-like tomographic weak lensing survey,
we adapt the general statistical approach developed by \citet{Baldauf2016}.This method incorporates the effects of non-negligible theoretical uncertainties in the covariance matrix, leading to a smooth suppression of modes where these uncertainties are larger. This method was recently applied for the case of unknown non-linear corrections in the matter and galaxy power spectra in  
\citet{Chudaykin_2021}.

This paper is organized as follows. 
In section \ref{sec:power} we review the theoretical modelling of the convergence power spectrum and its gaussian covariance matrix. Section \ref{sec:modcov} presents the main ingredients of a general proposal to include theoretical errors in a covariance matrix proposed in \citet{Baldauf2016}. In section \ref{sec:mitigation} we adapt this method to construct covariance matrices aimed at 
mitigating residual baryonic uncertainties using a set of hydrodynamical simulations and their best fits from a likelihood analysis that employed {\tt HMCode} to model baryonic effects.
In section \ref{sec:likelihood} we perform a simulated likelihood analysis for an LSST-like weak lensing survey with different covariance matrices and find that the extended covariance matrices in fact result in an increased accuracy (less biased inferred cosmological parameters) at the expense of a modest decrease in the precision (larger error bars). We discuss our findings in section \ref{sec:discussion} and present our conclusions in section \ref{sec:conclusion}.

\section{Convergence power spectrum and its gaussian covariance matrix}
\label{sec:power}
Here we are interested in the convergence angular power spectrum between two tomographic bins $i$ and $j$, $ C_{\kappa\kappa}^{ij}({\ell})$ given in the Limber approximation by:

\begin{equation}\label{Cl com lens efficiency}
    C_{\kappa\kappa}^{ij}({\ell}) =  \int^{\chi_h}_0 d\chi \frac{g^i(\chi) g^j(\chi)}{\chi^2} \, P_m \left(\frac{{\ell}}{\chi}, z(\chi)\right)
\end{equation}
where $\chi$ is the comoving radial distance between the observer and the object, the lens efficiency $g^i(\chi)$, in a flat cosmology, is written for source galaxies with redshift distribution $n^i(z)$ as:
\begin{equation}\label{lens efficiency}
    g^i(\chi) \equiv \frac{3\Omega_{m}H_0^2}{\,2c^2a(\chi)} \int^{\chi_h}_0 \, dz \, n^i(z) \frac{(\chi'(z) - \chi) \chi}{\chi'(z)} \,  \Theta(\chi'(z)-\chi) ,
\end{equation}
with $\Omega_\mathrm{m}$ the matter density parameter, $c$ the speed of light, $a(\chi)$ is the expansion scale factor as a function of $\chi$, $H_0$ the Hubble constant taken at the present day
and $\Theta(\chi'(z)-\chi)$ is the Heavyside step function. In this preliminary analysis we will not consider effects such as bias corrections to shear and intrinsic alignments. 

The Gaussian covariance of projected convergence power spectra can be expressed as  \citep{Hu:2003pt}
\begin{eqnarray}\label{Gaussian covariance}
    &\textrm{Cov}_{\textrm{G}}(C^{ij}_{\kappa \kappa}(\ell), \, C^{pq}_{\kappa \kappa}(\ell')) = \left< \Delta C^{ij}_{\kappa \kappa}(\ell) \, \Delta C^{pq}_{\kappa \kappa}(\ell') \right> = \\ \nonumber
    &\frac{2\pi \, \delta_{\ell \ell'}}{A \, \ell \, \Delta \ell} \left[\bar{C}^{ip}_{\kappa \kappa}(\ell)\bar{C}^{jq}_{\kappa \kappa}(\ell') + \bar{C}^{iq}_{\kappa \kappa}(\ell)\bar{C}^{jp}_{\kappa \kappa}(\ell')\right] ,
\end{eqnarray}
with 
\begin{equation}\label{C BAR}
    \bar{C}^{ij}_{\kappa \kappa}(\ell) = C^{ij}_{\kappa \kappa}(\ell) + \delta_{ij} \frac{(\sigma^i)^2}{n^i_A} ,
\end{equation}
where  $A$ is the angular survey area, $\Delta\ell$ is the angular bin width (as described in section \ref{sec:power}), $n^i_A$ is the area density of galaxies in redshift bin $i$ and $\sigma$ is the gaussian shape noise per component.
For LSST Y10, we adopt the requirements of \citet{2018arXiv180901669T} with a survey area $A$ of $14,300 \text{\, deg}^2$, shape noise of $\sigma^i = 0.26$ and $n^i_A = 5.4 \; \text{arcmin}^{-2}$ for all bins. Furthermore, we use a gravity-only model for the nonlinear 2-point function from \citet{Takahashi:2012em} in {\tt CosmoLike} \citep{Krause:2016jvl}, in order to generate the analytical Gaussian covariance matrix.

The covariance matrix has contributions from a Gaussian part and a non-Gaussian part composed of the connected 4-point (trispectrum) contributions and super-sample covariance \citep{Hu:2003pt,Krause2017,Barreira:2017fjz}. The Gaussian contribution is the dominant one as seen in a $\chi^2$ analysis for DES-Y3 set-up \citep{OliverCov} and for stage-IV experiments \citep{Barreira:2017fjz}.
As an initial test of our mitigation method we will be interested in incorporating errors from residual baryon effects in the Gaussian covariance matrix.

\section{Mitigating uncertainties with modified covariances}
\label{sec:modcov}
In this section we briefly review the strategy described in \citet{Baldauf2016} to model a general residual error as a gaussian random variable that can be marginalized over resulting in an additional contribution to the covariance matrix.

Let $\textbf{x}$ be the data vector, and $\textbf{t}$ the theoretical vector. The error vector $\textbf{e}$ being the residual between the data vector and its corresponding best-fit theory, with mean value $\bar{\textbf{e}}$. We assume $\textbf{e}$ to follow a Gaussian distribution
\begin{equation}\label{error dist}
    \mathrm{P_e} \propto \exp{ \Big[ -\frac{1}{2} (\textbf{e}-\bar{\textbf{e}})C_e^{-1} (\textbf{e}-\bar{\textbf{e}}) \Big]} ,
\end{equation}
with a covariance matrix $C_e$ given by
\begin{equation}
    C_e^{ab} = \left<e^a\,e^b\right> - \bar{e}^{\,a}\,\bar{e}^{\,b}.
\end{equation}
In this section, for simplicity, we will use $a$ and $b$ as the indexes for the angular bins.
We parametrize $ \left<e^a\,e^b\right>$ as
\begin{equation}
    \left<e^a\,e^b\right> \equiv E_a \, \rho_{ab} \, E_b,
\end{equation}

where we introduced a quantity we call the envelope $E_a=E(\ell_a)$ and assume that the correlation coefficient $\rho_{ab}$ is Gaussian and it depends only on the distance between two bins centered at $\ell_a$ and $\ell_b$. Thus,
\begin{equation}
    \rho_{ab} \equiv \exp\left[ -\frac{(\ell_a-\ell_b)^2}{2L^2} \right] .
\end{equation}

Hence, we can fully describe this mitigation approach by the smooth envelope $E(\ell)$ and the correlation length $L$, which specifies the minimal scale of variation of the theoretical model.

Assuming a Gaussian likelihood, we can include the theoretical error as 
\begin{equation} 
\mathcal{L}_e \propto \exp \bigg\{ -\frac{1}{2}\left[(\textbf{x}-\textbf{t}-\textbf{e})\,C_d^{-1}(\textbf{x}-\textbf{t}-\textbf{e}) + (\textbf{e}-\bar{\textbf{e}}) \, C_e^{-1}(\textbf{e}-\bar{\textbf{e}})\right] \bigg\} ,
\end{equation}
where $C_d$ is the usual data covariance matrix and $C_e$ the error covariance matrix.
We can marginalize the likelihood over the errors $\textbf{e}$ to obtain:
\begin{equation}\label{general likelihood}
    \mathcal{L} \propto \exp{\left[-\frac{1}{2}(\textbf{x} - \textbf{t} - \bar{\textbf{e}})C^{-1}(\textbf{x} - \textbf{t} - \bar{\textbf{e}})\right]} ,
\end{equation}
with the augmented covariance matrix $C$ given by
\begin{equation}\label{ERROR COV MATRIX}
    C = C_d + C_e.
\end{equation}
In the next section we will present our ansatz for the error covariance matrix in the case of uncertainties arising from the modelling of baryon physics.

\section{Modelling the theoretical error covariance}
\label{sec:mitigation}

In this section we use 13 hydrodynamical simulations to model the residual baryonic error from { \tt HMCode} on the
convergence angular power spectrum of an LSST-like survey with the introduction of a baryonic error covariance matrix, which we will denote in the following by $\mathrm{Cov^{Bar}}$. The 13 hydrodynamical simulations considered in this work are: Illustris \citep{Vogelsberger14}, Eagle \citep{Schaye15}, MassiveBlack-II \citep{Khandai15}, Horizon-AGN \citep{Dubois14}, and the 9 different baryonic scnearios from the OWLS simulation set \citep{Schaye10, vanDaalen11}.

We compute the tomographic convergence angular power spectrum from Eq.\eqref{Cl com lens efficiency} with an upper limit of $\ell_{\text{max}}\sim 3000$. Following the Dark Energy Science Collaboration (DESC) requirements for the LSST Y10 weak lensing  analysis \citep{2018arXiv180901669T1}, we consider 20 equally spaced logarithmic angular $\ell$ bins ranging from $20 \sim 3000$, for each tomographic spectra.

The matter power spectrum that enters Eq.\eqref{Cl com lens efficiency} includes baryons and nonlinear effects that we must account for.
One widely used method to take include these effects in the matter power spectrum is to use a phenomenological halo model based approach implemented in {\tt HMCode} \citep{Mead:2015yca}.
This variant of the halo model uses two physically motivated additional parameters: the halo bloating parameter, $\eta_0$, and the minimum halo concentration, $A$. Calibration with the {\tt Cosmic Emu} emulator obtained from the high resolution gravity only (G) N-body simulations Coyote suite \citep{Lawrence:2009uk} yields $A=3.13$ and $\eta_0 = 0.604$.
When varying the $A$ and $\eta_0$  parameters, one controls the halo-profile in a mass-dependent way that reproduces different feedback processes from various baryonic scenarios. 

The {\tt HMCode} can be used to reproduce the results from hydrodynamical simulations. However, the results from best-fit parameters arising from a Markov Chain Monte Carlo analysis show residual errors between the
{\tt HMCode}-generated power spectra and the power spectra from simulations \citep{Huang:2018wpy}.
We will mitigate these residual errors, modelling them as gaussian variables that can be marginalized, generating an augmented covariance matrix as reviewed in section \ref{sec:modcov}.

There are several hydrodynamical simulations that include the effects of baryons but they all depend on certain assumptions, such as the intensity of baryonic feedback processes.
We use the results from \citet{Huang:2018wpy} who studied the spread in the predictions of the 3D power spectrum from different hydrodynamical simulations to assess the residual errors.
We denote $P^\delta_{\textrm{hydro}}$ the 3D power spectrum output from a given hydrodynamical simulation. 

One difficulty in comparing different simulations is that they do not have the same input cosmology defined by the parameters that we denote $\textbf{p}_{co}$. 
In order to compare results for the same cosmology, one adopts the following definition for the baryonic power spectra, $P^\delta_{\textrm{hydro}}$:
\begin{equation}\label{sim pk}
    P^\delta_{\textrm{hydro}}(k,z|\textit{\textbf{p}}_{co}) = \frac{P^\delta_{\textrm{hydro, sim}}(k,z|\textit{\textbf{p}}_{co,sim})}{P_\delta^{\textrm{G,sim}}(k,z|\textit{\textbf{p}}_{co,sim})}  P^\delta_{\textrm{HMcode, G}}(k,z|\textit{\textbf{p}}_{co}) .
\end{equation}
where  $P^\delta_{\textrm{hydro, sim}}$ is the outcome from a given baryonic simulation at some cosmology $\textit{\textbf{p}}_{co,sim}$ and $P^\delta_{\textrm{G,sim}}$ denotes the corresponding G run. Finally, $P^\delta_{\textrm{HMcode, G}}(k,z|\textit{\textbf{p}}_{co})$, is the power spectrum calculated from the {\tt HMcode} calibrated by gravity only simulations.
Thus we are assuming that the baryonic physics contribution to the power spectrum
is independent of the input cosmologies $\textit{\textbf{p}}_{co,sim}$. This was shown to be a good approximation in \citet{vanDaalen:2019pst} by running hydro-simulations given the span of cosmology from WMAP 2009 \citep{WMAP9} to Planck 2015 \citep{Planck16}. \citet{Schneider20} also showed that ignoring the coupling between baryon and cosmology would be valid for future stage IV weak lensing experiments. 
We adopt the fiducial flat-$\Lambda$CDM cosmology shown in Table \ref{table:fid parameters}.

\begin{table}
\centering
\caption{Parameters of the  flat-$\Lambda$CDM cosmology adopted in this work. Massless neutrinos were assumed.}
\vspace{0.5cm}
\begin{tabular}{r|lr}
 
Parameter & Fiducial & Prior\\ 
\hline                               

$\Omega_m$    & 0.3156        & flat($0.2998 - 0.3314$) \\
$\sigma_8$    & 0.831         & flat($0.789 - 0.873$) \\
$h_0$         & 0.6727        & fixed                 \\
$\Omega_b$    & 0.0491685     & fixed                 \\
$n_s$         & 0.9645        & fixed                 \\
$w$           & -1.0          & fixed                 \\
$\tau$        & 0.08          & fixed                 
\label{table:fid parameters}
\end{tabular}
\end{table}

In order to compute the convergence angular power spectrum one needs to project the 3D power spectrum into different tomographic redshift bins. 
For the galaxy number distribution, we again DESC requirements for the LSST Y10 weak lensing  analysis \citep{2018arXiv180901669T1}. Hence, we use the following parametric form for the source redshift distribution $n(z)$:
\begin{equation} \label{parametric distribution}
    n(z) \propto z^2 \exp{\left[ -(z/z_0)^\alpha \right]} ,
\end{equation}
where we set $(z_0, \alpha) = (0.11,0.68)$ for Y10. Furthermore, the total number density of galaxies is normalized as $n^A= 27 \textrm{ arcmin}^{-2}$.

We take into account uncertainties in the photometric redshift measurements by considering a Gaussian probability distribution 
for a true redshift given a point measurement of a photometric redshift $z_{\textrm{phot}}$:
\begin{equation}\label{gaussian kernel}
    P(z_{\textrm{phot}}|z) \propto \exp{\left[-(z_{\textrm{phot}} - z)^2/2 \sigma_z^2 \right]} ,
\end{equation}
with a photometric redshift error of $\sigma_z = 0.05(1+z)$.
The redshift distribution in each photometric redshift bin $n_{i}(z)$  is then given by
\begin{equation*}
    n_{i}(z) = \int^{z_\textrm{phot}^{(i+1)}}_{z_\textrm{phot}^{(i)}} dz_{\textrm{phot}} \, n(z)\,  P(z_{\textrm{phot}}|z) ,
\end{equation*}
where the minimum redshift of the $i$-th tomographic bin, $z_\textrm{phot}^{(i)}$, is constructed such that each one contains an equal number density of galaxies, $n^i_A = 5.4 \; \text{arcmin}^{-2}$. Furthermore the number of galaxies per steradian in the i-th bin, $n_A^i$ is given by
\begin{equation*}
    n_A^i = \int_0^{\infty} dz \, n^i(z).
\end{equation*}

\begin{figure}
\centering
\includegraphics[width=0.47\textwidth]{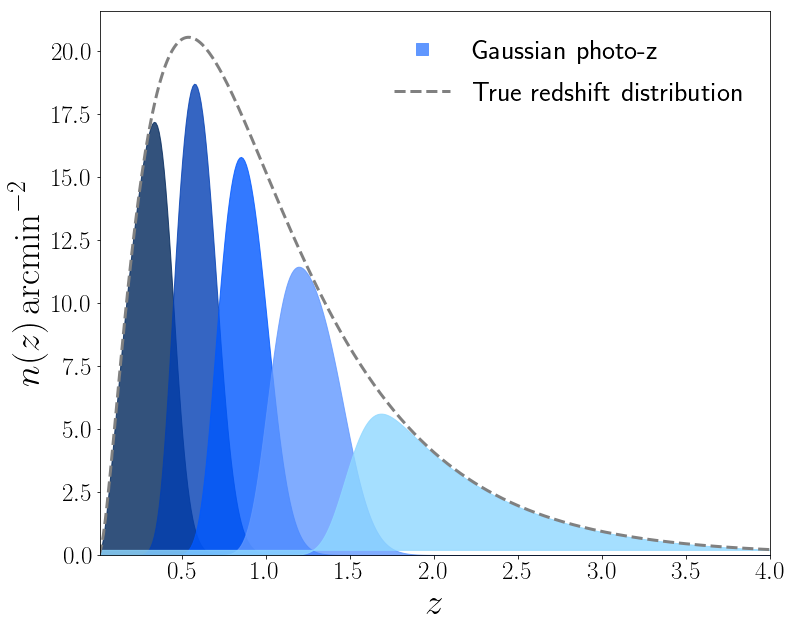}
\caption{The redshift distribution of source galaxies for LSST Y10 weak lensing measurements \citep{2018arXiv180901669T1}. Dashed line: The true underlying galaxy distribution following Eq. \eqref{parametric distribution} normalised to $n_A = 27 \; \text{arcmin}^{-2}$. Shaded areas: The redshift distribution of galaxies split into 5 tomographic bins normalised to $n^i_A = 5.4 \; \text{arcmin}^{-2}$. The shades of blue are darker for lower redshifts and lighter for higher redshifts.}
\label{galaxy distribution}
\end{figure}

The resulting five $n^{i}(z)$ tomographic distributions for the LSST source samples are shown in Fig. \ref{galaxy distribution}. By construction, the sum of the individual distributions equals the total $\sum_i n^i_A \equiv n^A$. 

With the galaxy redshift distributions in the five tomographic bins we can proceed to model the residual errors of baryonic effects on the convergence angular power spectrum.
We compare the convergence angular power spectra as obtained from a given hydrodynamical simulation with the best-fit {\tt HMCode} results for that particular simulation. An example of these residual errors for 13 different simulations is shown in Figure \ref{Figure:Envelopes}, for the auto-correlations of redshift bins 0 and 4 in the same angular binning and scale-cut as the data-vector. One can see that the spread of the errors decreases at higher redshifts, where baryonic effects are less important.

In this work, we use the best-fit models generated by \citet{Huang:2018wpy}. These models are characterized by the best-fit values of $A$  and $\eta_0$ and were obtained from Markov Chain Monte Carlo (MCMC) runs, fitting the {\tt HMCode} baryonic parameters to the 2D convergence power spectra of the hydrodynamical simulations (considering the 3D power spectrum of Eq. \eqref{sim pk}).
We will now use these results to model the two ingredients that enter the additional covariance matrix due to the marginalization of the baryonic residual error: the envelope and the correlation. 

\subsection{Modelling the envelope}
Based on the residual errors for the angular power spectra and on the assumption that the true angular power spectrum spectra (i. e., the one directly obtained from observations) lies among the range of the hydrodynamical models, we decided to test three different parametrizations for the envelope shown in Figure \ref{Figure:Envelopes} that we call the Mirror, 2Mirror and the Variance envelopes.
	
The Mirror envelope is a conservative definition. It takes the most extreme deviations of the HMcode best-fit models and mirrors them about the x-axis, hence the name Mirror envelope. This approach overestimates the error amplitude, but guarantees that we are taking all the possible deviations the baryonic error may present. Also, this definition ensures the residual error, $\textbf{e}$, to have zero mean, $\bar{\textbf{e}}=0$, which leads to the additional baryonic covariance matrix:
\begin{equation}\label{mirror-covariance}
	    \mathrm{Cov_{Bar}}(C_{\kappa\kappa}^{ij}(\ell), C_{\kappa\kappa}^{pq}(\ell')) = E^{ij}_{\textrm{Mirror}}(\ell) \,\, \rho^{ij, pq}(\ell,\ell') \,\, E^{pq}_{\textrm{Mirror}}(\ell')
\end{equation}
with 
	\begin{equation}\label{mirror env}
	    E^{ij}_{\textrm{Mirror}}(\ell) \equiv C_{\kappa\kappa}^{ij}(\ell) \, \left|1-\frac{C^{ij}_{\textrm{HMcode model}}(\ell)}{C^{ij}_{\textrm{Sim model}}(\ell)}\right|_{\text{model=max}} ,
	\end{equation}
where $C_{\kappa\kappa}^{ij}(\ell)$ on the right-hand-side follows a gravity only model, just like the one used to compute the Gaussian covariance in Eq. \eqref{Gaussian covariance}.
	 we choose the boundaries of the envelope to be at the model that presents the maximum deviation at that $\ell$. The absolute value makes it explicit that the Mirror envelope is a symmetric function on the $\ell$-axis.
An even more conservative envelope, used to stress-test our approach, is the 2Mirror envelope which consists in simply doubling the Mirror envelope.

The Variance envelope, on the other hand, is a less conservative approach
which defines the envelope as the variance of the random vector $\textbf{e}$. 
In this approach, we interpret the residual error from each hydrosimulation 
as a realization of the random variable $\textbf{e}$. 
Hence, we simply take the variance between the thirteen error curves and define it as our envelope, as follows:
\begin{equation}\label{sigma-envelope}
	    E^{ij}_{\textrm{Var}}(\ell) = C_{\kappa\kappa}^{ij}(\ell)\sqrt{\frac{1}{N} \sum^N_{\textrm{model}} \left[1-\frac{C^{ij}_{\textrm{HMcode model}}(\ell)}{C^{ij}_{\textrm{Sims model}}(\ell)} -\bar{e}^{\,ij}(\ell) \right]} ,
\end{equation}	
where $N=13$ stands for the total number baryonic models being considered here.
In contrast with the mirror envelope, this definition does not impose a symmetric envelope; in other words, the Variance approach admits a non-zero mean value for the theoretical error, $\bar{e}^{\,ij}(\ell) \equiv \left< 1 - \frac{C_{\textrm{HMcode}}^{ij}(\ell)}{C_{\textrm{Sims}}^{ij}(\ell)} \right>$ 
and the baryonic error contribution to the covariance matrix has to be changed accordingly:

	\begin{align}
	    \mathrm{Cov_{Bar}}(C_{\kappa\kappa}^{ij}(\ell), C_{\kappa\kappa}^{pq}(\ell')) &= E^{ij}_{\textrm{Var}}(\ell) \,\, \rho^{ij, pq}(\ell,\ell') \,\, E^{pq}_{\textrm{Var}}(\ell') \label{sigma covariance}
	    \nonumber \\ &- \, \bar{e}^{\, ij}(\ell) \, \, \bar{e}^{\, pq}(\ell')
	\end{align}

Figure \ref{Figure:Envelopes} shows the different envelopes for two redshift bins: the first and the last ones. As expected, larger redshifts result in larger physical scales for the same angular scale leading to a decrease in the baryonic effects for a given angular scale. Notice that, as opposed to the mirror model, the Variance envelope underestimates the covariance amplitude for the most extreme scenarios. For instance, the Illustris simulation residual errors (red line) are left outside of the Variance envelope for $\ell \gtrsim 100$.

\subsection{Modelling the correlation}
The last ingredient to model is the correlation coefficient $\rho^{ij, pq}(\ell,\ell')$ that relates different redshift bins and Fourier modes of the error covariance.
We adopt the ansatz
\begin{equation}\label{correlation}
    \rho^{ij, pq}(\ell,\ell') \equiv R^{ij,pq} \exp{\left[ -(\ell - \ell')^2/2 L^{ij} L^{pq} \right]},
\end{equation}
that separates the redshift bin correlations $R^{ij,pq}$ from the correlation of Fourier modes. 

In this work we will model the effect on the covariance within the same redshift bin pairs, neglecting cross-covariances induced by baryonic effects in different redshift bin pairs, \ie we assume:
\begin{equation}\label{redshift correlation}
    R^{ij,pq} = \delta^{ip} \delta^{jq}.
\end{equation}
With this major assumption, we are including tomographic power spectra that can fluctuate independently from other tomographic pairs as possible baryonic models. We will show that this ansatz is sufficient to mitigate the baryonic uncertainties.

With a diagonal $R^{ij,pq}$, the only parameter left to fully define the theoretical error covariance is the correlation length of the baryonic errors, $L^{ij}$.
We adopt
\begin{equation}\label{mono scale}
    L^{ij} = k_{\textrm{halo}}  \left< \chi \right>^{ij} = k_{\textrm{halo}} \frac{\int d\chi \,\, \chi \,  g^i(\chi) g^j(\chi)}{\int d\chi \,\, g^i(\chi) g^j(\chi)},
\end{equation}
with $k_{\text{halo}} = 1.0 \, h /\textrm{Mpc}$ being a typical halo scale for $\rho_{\textrm{virial}}=\rho_{200}$ and $\mathrm{M}_{200} \approx 10^{13.5} \mathrm{M}_{\odot}$. The chosen halo mass input, $\mathrm{M}_{200} \approx 10^{13.5} \mathrm{M}_{\odot}$, was motivated by \cite{Takada_2007}. In their Figure 3, they show that, at non-linear scales, an expressive fraction of the 1-halo term contributions for the lensing effects comes from halos with masses of $ \approx 10^{13.5} \mathrm{M}_{\odot}$.
The calculated values of $L^{ij}$ using Eq. \eqref{mono scale} are shown in Table \ref{table:mono scale}.

\subsection{Full covariance}
Finally, the full covariance is given by
\begin{align}
    \mathrm{Cov}(C_{\kappa\kappa}^{ij}({\ell}), C_{\kappa\kappa}^{pq}({\ell'})) &= \mathrm{Cov_{G}}(C_{\kappa\kappa}^{ij}({\ell}), C_{\kappa\kappa}^{pq}({\ell'})) \, \delta_{\ell}^{\ell'}  \label{gaussian+baryonic covariance} \nonumber \\ &+\, \mathrm{Cov_{Bar}}(C_{\kappa\kappa}^{ij}({\ell}), C_{\kappa\kappa}^{pq}({\ell'}))    ,
\end{align}
where the Gaussian covariance matrix is given by Eq. (\ref{Gaussian covariance}) and
is analytically generated using {\tt CosmoLike} \citep{Krause:2016jvl} with the LSST survey characteristics already discussed. 

\begin{table}
\centering
\caption{Evaluated tomographic values of the characteristic $\ell$-scale for residual baryonic errors, $L^{ij}$,  calculated in Eq. \eqref{mono scale}.}\label{table:mono scale} 
\vspace{0.1cm}
\begin{tabular}{|l|*{5}{c|}}
      \hline
      \diagbox[width=\dimexpr \textwidth/100+4\tabcolsep\relax, height=.6cm]{ i }{ j }
                   & 0 & 1 & 2 & 3 & 4 \\
\hline                            
0 & $491$ & $600$ & $637$ & $653$& $665$ \\
\hline
1 & -- & $806$ & $912$ & $959$ & $987$\\
\hline
2 & -- & -- & $1120$ & $1241$ & $1308$\\
\hline
3 & -- & -- & -- & $1483$ & $1655$\\
\hline
4 & -- & -- & -- & -- & $2112$ \\
\hline
\end{tabular}

\end{table}

It is important to mention that shape noise starts to dominate the gaussian covariance matrix, that is $(\sigma^i)^2/2n^i_A > C^{ii}_{\kappa\kappa}(\ell)$ in Eq. \eqref{C BAR}, for $\ell \gtrsim 600$, in the last redshift bin. For closer redshift bins, the shape noise is dominating for even smaller values of $\ell$.

In Figure \ref{Figure:covariance} we show the fractional difference between the Gaussian covariance matrix and the augmented covariance matrix (Gaussian plus baryonic theoretical errors). For large $\ell$ (small scales), the theoretical errors term has larger relative values and dominates the uncertainties.   
In the following section, we study the impact of the augmented covariance matrix on parameter estimation using a simulated likelihood analysis.  

\section{Likelihood analysis}
\label{sec:likelihood}
In this section we present our analysis choices used to assess the effectiveness of the proposed mitigation approach.
In general lines, the analysis methodology consists in the following steps:
\begin{mylist}
    \item[(\rom{1})] Take the lensing spectra predicted by one of the four hydrodynamical simulations (Eagle, Illustris, MB-II and Horizon-AGN) as mock data for our fiducial cosmology. These four simulations are representatives of the different baryonic effects. \label{step1}
    \item[(\rom{2})] Use the nested sampling algorithm {\tt Multinest} \citep{Feroz:2008xx} to fit the mock data to the {\tt HMCode} model by varying two cosmological ($\Omega_m$, $\sigma_8$) and two nuisance parameters ($A$, $\eta_0$) and determine the statistical errors from the Gaussian covariance on the cosmological parameters. 
This analysis does not include the residual effects of the model.  \label{step2}
    \item[(\rom{3})]  Determine the {\tt HMCode} bias with respect to the input cosmological parameters before the mitigation technique is applied. \label{step3}
    \item[(\rom{4})]  Use {\tt Multinest} to fit the mock data again to a model with those same two cosmological and two nuisance parameters but with the augmented covariance matrix
proposed in the previous section.   \label{step4}
    \item[(\rom{5})] Determine the final residual bias as the difference between this second fit and the true values of the cosmological parameters used in the mock data (specified in Table \ref{table:fid parameters}).  \label{step5}
    \item[(\rom{6})] Compare the statistical degradation of the methods from the size of the error bars in both cases.  \label{step6}
\end{mylist}

In our analysis, we consider three sets of results for baryonic mitigation: {\tt HMCode} + Gaussian covariance matrix,  {\tt HMCode} + Mirror covariance mitigation and 
{\tt HMCode} + Variance covariance mitigation. The 2Mirror covariance mitigation is used as a stress test of the method for the Illustris and MassiveBlack-II simulations.

\begin{figure*}
\centering
\includegraphics[width=1.0\textwidth]{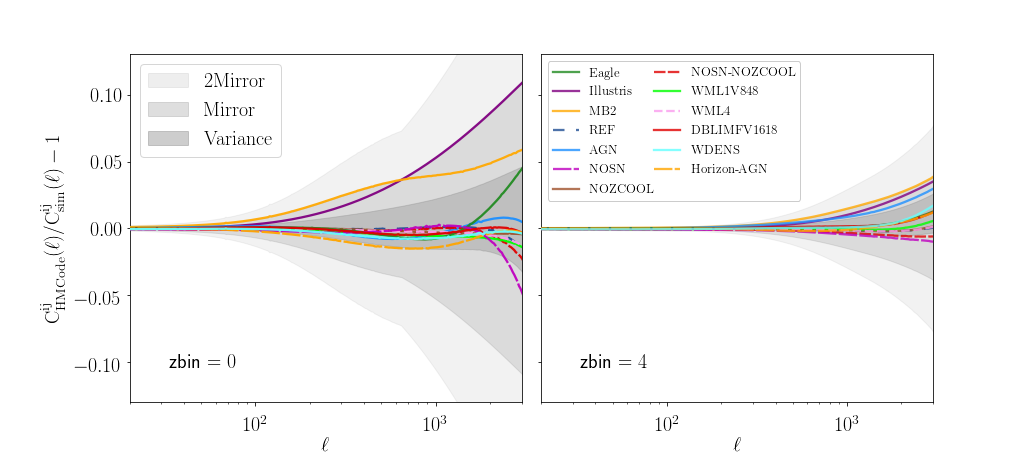}
\caption{The figures show the envelope's behaviour, at redshift bins $i,j = 0, 0$ (low-redshift, on the left) and $i,j = 4, 4$ (high-redshift, on the right), with respect to the set of 13 baryonic models (solid lines). The lighter shaded area presents the 2Mirror envelope definition; which doubles the size of the most extreme scenarios and reflect them into the x-axis, ensuring a zero mean to the error variable. The intermediate grey area shows the coverage of the Mirror envelope. In this definition, the covariance amplitude is assumed to follow the size of the most extreme scenario. Finally, the darkest area shows the Variance envelope. For this definition, we take the standard deviation between all scenarios as the amplitude.}
\label{Figure:Envelopes}
\end{figure*}

\begin{figure*}
\centering
\includegraphics[width=0.46\textwidth]{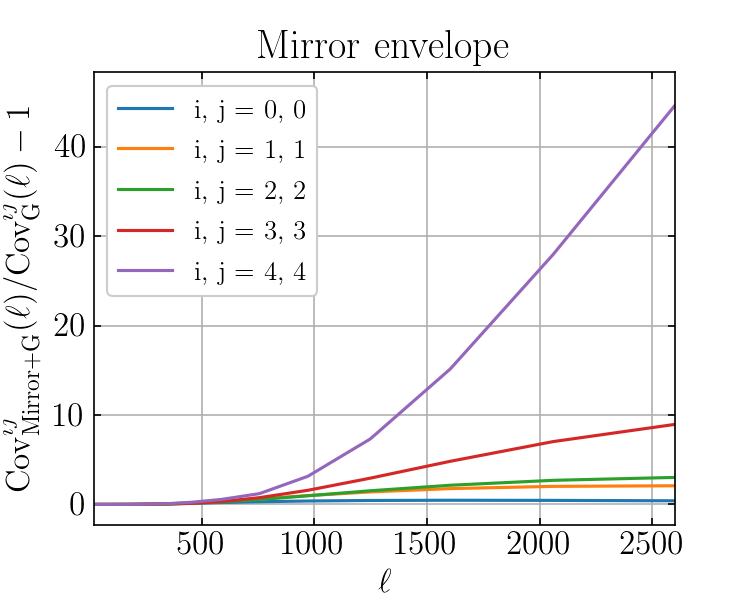}
\includegraphics[width=0.46\textwidth]{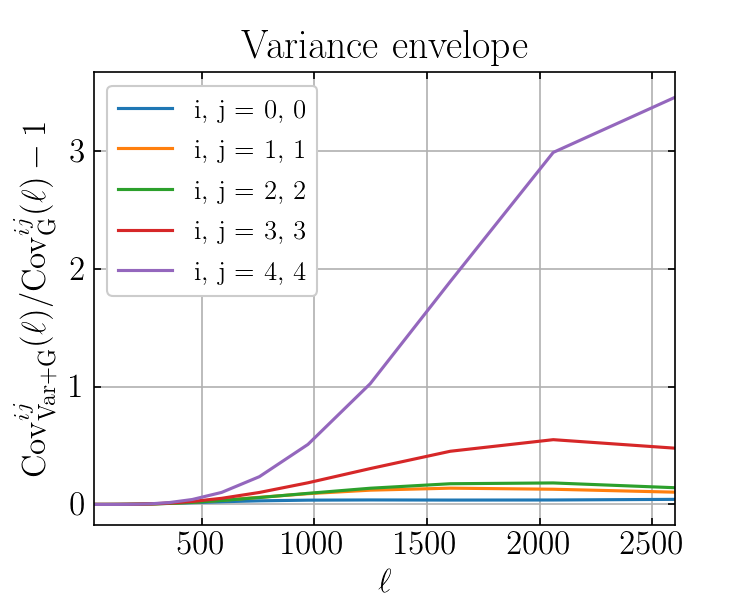}
\caption{Fractional change in the diagonal elements of the covariance matrix  ($\ell=\ell'$) due to the inclusion the the baryonic terms from the Mirror (left) and Variance (right) envelopes.}
\label{Figure:covariance}
\end{figure*}

We assume a Gaussian likelihood for the tomographic two-point measurements.
The model predictions are computed with {\tt HMCode} and the
final posterior distribution on cosmological parameters is obtained with  {\tt Multinest}, implemented in {\tt CosmoSIS} \footnote{\url{https://bitbucket.org/joezuntz/cosmosis/wiki/Home}} \citep{Zuntz:2014csq}. We use $n_{\textrm{live}}=100$ live points, efficiency  of $0.01$ and tolerance of $0.01$.
We sample over the parameters $\{ \Omega_m, \sigma_8,  A, \eta_0 \}$ since we want to concentrate on the cosmological parameters mostly affected by baryonic effects.

The results of the posterior distributions for the four different covariance matrices are shown in Figures \ref{fig:2dB} and \ref{fig:2dD} for the Illustris and MassiveBlack-II simulations. 
The dashed lines show the input cosmological parameters $(\Omega_m, \sigma_8)$ together with the HMCode parameters $(A, \eta_0)$ determined from the best fit analysis for a given simulation.
For the Illustris simulations one can notice a significant decrease in the bias for the cosmological parameters whereas for the MassiveBlack-II simulations there is a less significant improvement. 
This is probably due to the fact that Illustris has a stronger baryonic feedback than MassiveBlack-II. 
It is also interesting to notice the effect of "saturation" of the theoretical covariance matrix by comparing the results from the Mirror to the 2Mirror matrices: by becoming very conservative one stops losing statistical power and hence the areas of the ellipses do not change significantly. This is due to the fact that the affected modes are already suppressed and further suppression does not remove information. 
However, it is important to point out that there is a slight difference between different choices of the covariance matrix. This can be the most easily seen looking at the 2D posteriors for degenerate parameters, such as $\Omega_m-\sigma_8$ panel in Figure~\ref{fig:2dB}. In this case the true cosmology for the most aggressive $1\sigma$ envelope is slightly outside the $1\sigma$ contour. On the other hand, using the more conservative Mirror covariance leads to unbiased results for both cosmological parameters as well as for the best constrained principal component. This is important to keep in mind, particularly for combination with external data which have different degeneracy directions.

The results for the  1D marginalized $68 \%$ error bars for the cosmological parameters $(\Omega_m,\sigma_8)$ are shown in Figure \ref{fig:1dA} for the four simulations using three different covariance matrices for the analyses (four for Illustris and MassiveBlack-II simulations). One can see that by using improved covariance matrices modelling baryonic uncertainties can help in reducing the bias on the determination of cosmological parameters at a modest increase of the uncertainties. 
\section{Discussion}
\label{sec:discussion}

Our final results are summarized in Table \ref{main results}, where we show 
the $68\%$ error bars on the parameters $\Omega_m$, $\sigma_8$, $A$ and $\eta_0$ for four simulations and different covariance matrices and the amount of bias in the cosmological parameters $\Omega_m$ and $\sigma_8$ measured in units of the standard deviation.
For Illustris and MassiveBlack-II, representatives of extreme cases of baryonic parameters, we also present results with the very conservative case of the 2Mirror envelope. We can think of these augmented covariances from the Mirror and Variance envelopes acting in the data vector as a soft scale-cut. They gradually reduce the weight of a data point for the overall analysis as we move to scales with larger theoretical uncertainties. 

\begin{figure}
    \centering
    \begin{subfigure}{0.47\textwidth}
        \includegraphics[width=\textwidth]{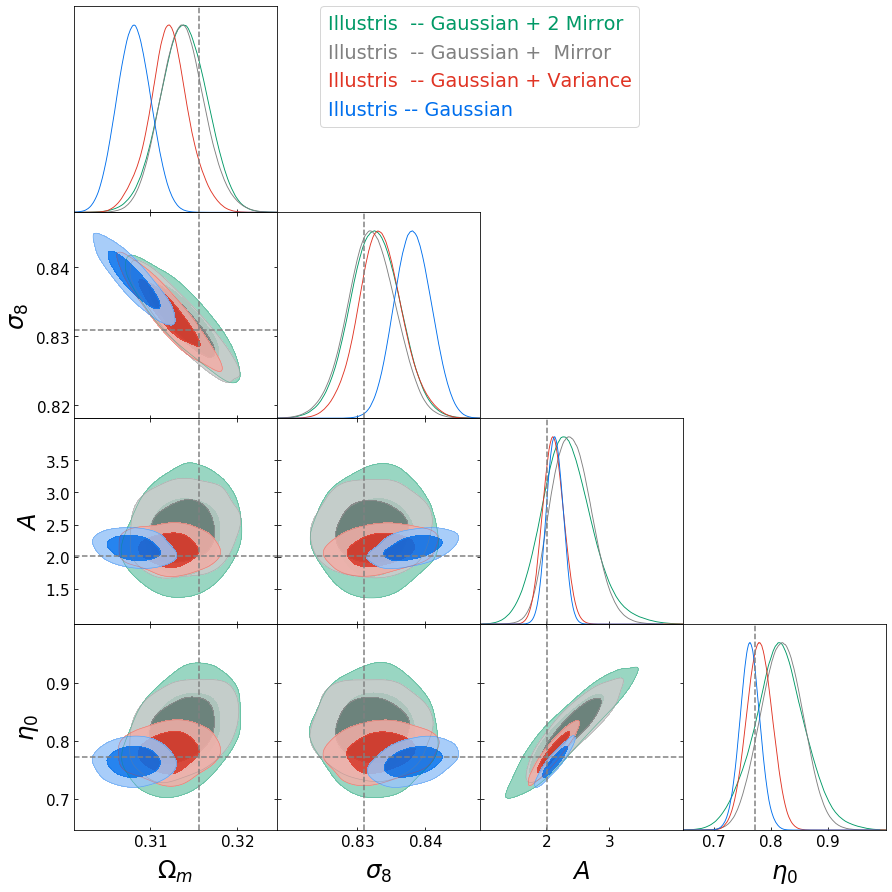}
        \caption{Posterior distributions for an Illustris data vector.}
        \label{fig:2dB}

\end{subfigure}\\
\begin{subfigure}{0.47\textwidth}
        \includegraphics[width=\textwidth]{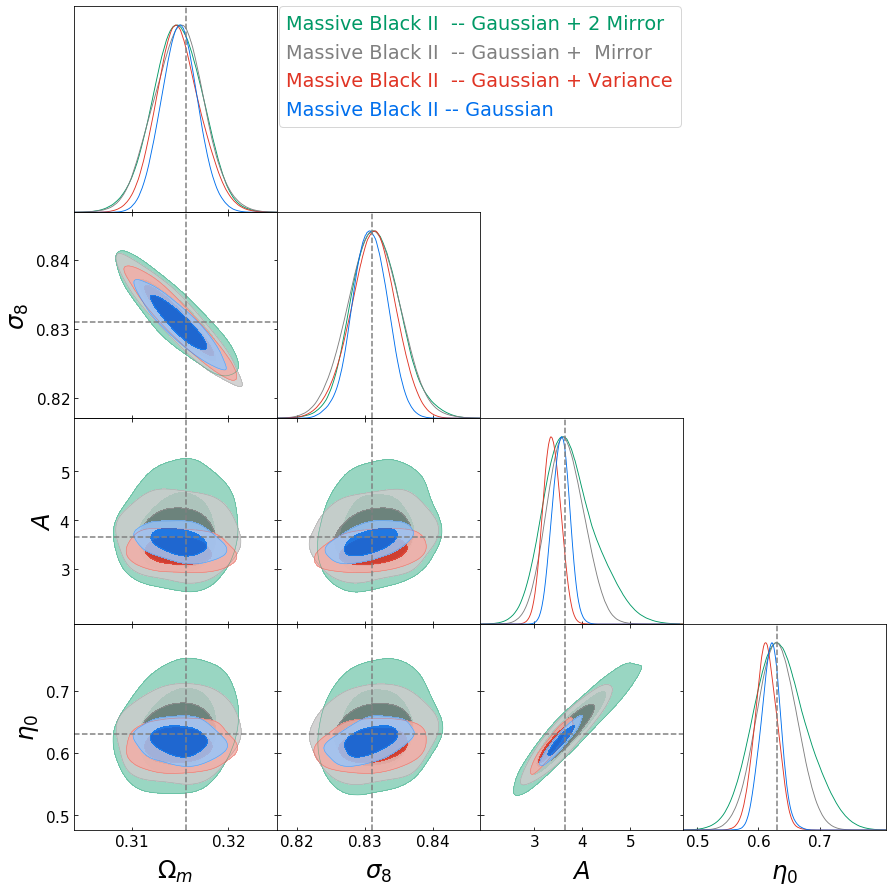}
            \caption{Posterior distributions for a MassiveBlack-II data vector.}
            \label{fig:2dD}
\end{subfigure}
\caption{Comparison of the posterior distributions for the cosmological $(\Omega_m, \sigma_8)$ and nuisance
parameters $(A, \eta_0)$ among different mitigation covariances. The colors represent different covariance matrix model: gaussian contributions only (blue),
Mirror envelope (grey), 2Mirror (green) and Variance envelope (red) taking the Illustris (left) and MassiveBlack-II (right) simulations as the input data vector.
The dashed lines show the input cosmological parameters $(\Omega_m, \sigma_8)$ together with the HMCode parameters $(A, \eta_0)$ determined from the best fit analysis for the simulations.}

\end{figure}

\begin{figure*}
    \centering
    \includegraphics[width=0.8\textwidth]{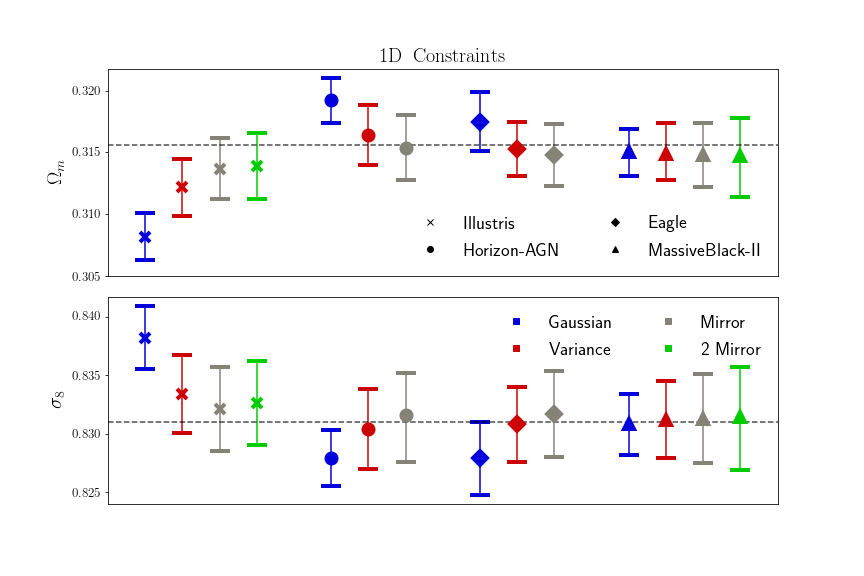}
    \caption{Results for the $68 \%$ error bars for the cosmological parameters $\Omega_m$ and $\sigma_8$ for different data vectors from 4 hydrodynamical simulatiosn using three different covariance matrices (4 in the case of Illustris and MassiveBlackII) in a nested sampling analysis of the posterior. Dashed lines are the input parameters.}
    \label{fig:1dA}
\end{figure*}

\begin{table*}
    \centering
    \caption{Summary of the results from simulated likelihood accuracy tests. In columns 3-6 we give the best-fit posterior values as well as the $68\%$ confidence interval for two $\Lambda \mathrm{CDM}$ cosmological and nuisance parameters of the {\tt HMCode} ($\Omega_m$ and $\sigma_8$; $A$ and $\eta_0$). Columns 7-8 presents the offset between the best-fit values and the fiducial ones. This offset is quantified in terms of the $68\%$ confidence interval, i.e., the $1\sigma$ interval size of that constraint.}\label{main results}
    \vspace{0.5cm}
    
    \resizebox{\textwidth}{!}{\begin{tabular}{|l|c|cccc|cc|}
    \hline
     \multicolumn{1}{|c|}{\textbf{Baryonic}} & \multicolumn{1}{c|}{\textbf{Covariance}} & \multicolumn{4}{c|}{\textbf{68\% limits}} & \multicolumn{2}{c|}{\textbf{Bias}}\\
    \cline{3-8}
    \multicolumn{1}{|c|}{\textbf{data vector}} & \textbf{Matrix} & $\bm{\Omega_m}$ & $\bm{\sigma_8}$ & $\bm{\eta_0}$ & $\bm{A}$ & $\bm{\Omega_m}$ & $\bm{\sigma_8}$ \\
    \hline
    {Horizon-AGN} & Gaussian & $0.3192\pm 0.0018$ & $0.8267\pm 0.0024$  & $0.601^{+0.015}_{-0.018}$ & $2.54^{+0.13}_{-0.16}$ & $2.00 \, \sigma$ & $1.79 \, \sigma$ \\
    & Gauss. + Mirror   &  $0.3154\pm 0.0026$  & $0.8316^{+0.0040}_{-0.0036}$  & $0.587\pm 0.033$ & $2.51\pm 0.32$ & $0.08 \, \sigma$ & $0.16 \sigma$  \\
    & Gauss. + Variance &  $0.3164 \pm 0.0024$ & $0.8304 \pm 0.0034$  & $0.600\pm 0.021$ & $2.64\pm 0.19$ & $0.33 \, \sigma$ & $0.18 \, \sigma$ \\
    \hline
        {ILLUSTRIS} & Gaussian & $0.3082\pm 0.0019$  & $0.8382\pm 0.0027$  & $0.764\pm 0.017$ & $2.14\pm 0.13$ & $3.89 \, \sigma$ & $2.66 \, \sigma$\\
    & Gauss. + 2\textrm{Mirror}    &  $0.3139\pm 0.0027$  & $0.8326\pm 0.0036$ & $0.816\pm 0.046$ & $2.33^{+0.36}_{-0.42}$ & $0.63 \, \sigma$ & $0.45 \, \sigma$  \\
    & Gauss. + Mirror  &  $0.3137\pm 0.0025$  & $0.8321\pm 0.0036$ & $0.819\pm 0.036$ & $2.40^{+0.29}_{-0.34}$ & $0.76 \, \sigma$ & $0.31 \, \sigma$  \\
    & Gauss. + Variance &  $0.3122\pm 0.0023$  & $0.8334\pm 0.0033$  & $0.780\pm 0.022$ & $2.11\pm 0.16$ & $1.50 \, \sigma$ & $0.73 \, \sigma$ \\
    \hline
        {EAGLE} & Gaussian & $0.3175\pm 0.0024$  & $0.8279\pm 0.0031$  & $0.570^{+0.016}_{-0.015}$ & $2.52^{+0.13}_{-0.15}$ & $0.79 \, \sigma$ & $1.00\, \sigma$\\
    & Gauss. + Mirror  &  $0.3148\pm 0.0025$  & $0.8317\pm 0.0037$  & $0.569\pm 0.034$ & $2.56\pm 0.33$ & $0.32 \, \sigma$ & $0.19 \, \sigma$  \\
    & Gauss. + Variance &  $0.3153\pm 0.0022$  & $0.8308\pm 0.0032$  & $0.572\pm 0.020$ & $2.63^{+0.16}_{-0.18}$ & $0.13 \, \sigma$ & $0.06 \, \sigma$ \\
    \hline
    {MassiveBlack-II} & Gaussian & $0.3150\pm 0.0019$   & $0.8308\pm 0.0026$  & $0.620\pm 0.016$ & $3.56\pm 0.18$ & $0.32 \, \sigma $ & $0.07 \, \sigma$\\
         & Gauss. + 2\textrm{Mirror} &  $0.3147^{+0.0033}_{-0.0031}$  & $0.8314^{+0.0045}_{-0.0043}$ & $0.631^{+0.057}_{-0.052}$ & $3.63^{+0.72}_{-0.62}$ & $0.28 \, \sigma$ & $0.09 \, \sigma$  \\
    & Gauss. + Mirror    &  $ 0.3148\pm 0.0026$  & $0.8313\pm 0.0038$  & $0.630\pm 0.032$ & $3.66\pm 0.39$ & $0.31 \, \sigma$ & $0.08 \, \sigma$  \\
    & Gauss. + Variance &  $0.3149^{+0.0021}_{-0.0025}$  & $0.8312\pm 0.0033$  & $0.612\pm 0.018$ & $3.37 \pm 0.19$ & $0.30 \, \sigma$ & $0.06 \sigma$ \\
    \hline
    \end{tabular}\label{table:1d constraints}}
\end{table*}

Whereas the Mirror method performs a more conservative cut by accounting for unrealistically strong feedback models in its error amplitude, the Variance envelope considers the uncertainties on modeling more realistic AGN suppression leading to a softer cut. Figures \ref{fig:2dB}, \ref{fig:2dD} and \ref{fig:1dA} show the 2D and 1D constraints obtained through these two approaches. We now discuss our results for the different simulations according to the strength of AGN feedback.
\\[10pt]

\centerline{\textbf{Weak AGN model}}

We begin by discussing the performance of the {\tt HMCode}-only analysis, without the mitigation of its residuals (named as `Gaussian'). For the MassiveBlack-II (MB-II) data vector, the halo-bloating ($\eta_0$) and concentration parameters ($ A $) alone successfully recover the true cosmology. Even with only two free cosmological parameters, which may 
increase the bias since the other parameters are kept fixed, the best-fit values for both $\Omega_m$ and $\sigma_8$ are below the $\sim 0.4 \sigma$ offset shift. This result is not unexpected if we recall MB-II's response function shown in Fig. 1 from \cite{Huang:2018wpy}. Consequently, applying the 2Mirror, Mirror and Variance methods to this well-modelled scenario does not significantly affect the residual biases, which remains below the $\sim 0.4 \sigma $ deviation.
\\[10pt]

\centerline{\textbf{Strong AGN model}}

Fig. \ref{fig:1dA} shows the evolution of the marginalized bias, for mock data based on Illustris, over increasingly conservative approaches (from left to right). The blue bars represent the constraint obtained when relying only on the {\tt HMCode} mitigation parameters. When ignoring A and $\eta_0$ limitations on fitting complex dynamics, the residual bias goes highly above the $2\sigma$ deviation for both $\Omega_m$ and $\sigma_8$, as depicted in Table \ref{table:1d constraints}. However, studies on the {\tt HMCode} residuals obtained a different result from ours. \cite{Huang:2018wpy} obtained that, after marginalizing over 6 $w\mathrm{CDM}$ cosmological parameters, the halo-based model effectively mitigates the bias impact to less than $0.5\sigma$ for the $\Omega_m$ and $\sigma_8$ 1D posteriors. The discrepancy between our results is likely due to our bias analysis being naturally overestimated by the limited parameter space, especially on the $w_0$ and $w_a$ parameters with a strong correlation with $\Omega_m$ and $\sigma_8$\footnote{Since $w_0$ and $w_a$ are strongly correlated with $\Omega_m$ and $\sigma_8$, as we can see from Huang et al. Fig. 4, keeping them fixed during the likelihood analyses increases the information on the matter parameters posterior distributions. That information gain leads to a tighter constraint on $\Omega_m$ and $\sigma_8$ and, thus, on the overall bias.}.

The Variance covariance matrix (red bars) drastically reduces the {\tt HMCode}'s offset of $\Omega_m$ from $3.9\sigma$ to $1.5\sigma$, and $\sigma_8$ from $2.7\sigma$ to $0.7\sigma$, as shown in Table \ref{table:1d constraints}. If focusing on the more conservative method (grey bars) one sees that, in this extreme AGN model, the larger covariance amplitude pays itself in the bias mitigation: the offset of both cosmological parameters is less than the marginalized statistical uncertainty.
\\[10 pt]

\centerline{\textbf{Intermediate AGN models}}

Fig. \ref{fig:1dA} shows the marginal 1D distributions for the analysis based on AGN models that are not as underestimated as MB-II and not as unrealistically strong as Illustris. For the Horizon-AGN, both Mirror and Variance covariances are effective in reducing {\tt HMCode}'s residual error below the $0.4\sigma$ shift for our setup. The mirror is more successful in reducing the bias of both parameters: reaching a $0.1\sigma$ shift for $\Omega_m$ and $0.2\sigma$ deviation for $\sigma_8$. Compared with the Variance envelope, the Mirror approach degrades the 1D error of $\sigma_8$ by $ 17\%$ to gain $ 10\%$ on the accuracy. For $\Omega_m$, the loss in statistical power, $8\%$, compared to Variance statistics, is compensated by a $ 76\%$ accuracy improvement. 

We can understand Variance and Mirror's different performances with Horizon-AGN by recalling the shape of their covariances amplitudes from the left panel of Fig. \ref{Figure:Envelopes}. We can see that, for tomographic bins $i=0$ and $j=0$, the model's physics becomes underestimated as we move to beyond $\ell=1000$. On the other hand, the top left panel shows that the Mirror approach's amplitude has no problems with covering the same physics, leading to higher effective accuracy.

For the EAGLE based analysis, {\tt HMCode} fitting approach alone is effective enough to keep the bias within the $1 \sigma$ statistical uncertainty for one marginalized cosmological parameter. Furthermore, our residual mitigation methods improve the {\tt HMCode} model accuracy on both $\Omega_m$ and $\sigma_8$ from $0.8\sigma$ and $1.0\sigma$, respectively, to less than $0.2\sigma$ ($0.4\sigma$) for the Variance (Mirror) approach. 

Compared to the Gaussian method, in which we only rely on the {\tt HMCode} mitigation, the Mirror method increases the error bar by $4\%$ and $19\%$ for $\Omega_m$ and $\sigma_8$, respectively. Whereas for our less conservative mitigation covariance, the Variance method, the constraint on $\sigma_8$ degrades just by $3\%$ and it \textit{shrinks} for $\Omega_m$ by $8\%$. The first thing we can comment about these results is that both Mirror and Variance methods accuracy overcompensates the loss in precision, which means that they can extract more information from the likelihood analysis for the EAGLE scenario. Finally, we can see the gain of $8\%$ in statistical power in $\Omega_m$, from the $1\sigma$ analysis, even though we would expect the modified covariance matrix to degrade the cosmological constraints, as a consequence of the non-linear relation between the data covariance matrix and the posterior 1D distribution on the cosmological parameters.
\\[8pt]

To summarize, our modified covariance models (Mirror and Variance) are successful in improving the {\tt HMCode} information gain in the cosmic-shear likelihood analysis. For the "strong" and "intermediate" baryonic scenarios (Horizon-AGN, ILLUSTRIS, and EAGLE), the accuracy refinement of our method dramatically outweighs the loss in statistical power of the tested cosmological parameters (compared to the Gaussian analysis). The MassiveBlack-II scenario is the only exception in which our mitigation method does not seem to be necessary because it is already well-modeled by the {\tt HMCode} free parameters alone.

\section{Conclusions}
\label{sec:conclusion}
Baryonic physics can significantly affect the theoretical modelling of the matter power spectrum in the small-scale regime. Therefore mitigation methods have to be developed and tested to properly take this source of uncertainty into account.
In this work we focused on the mitigation of the baryonic effects using as an example the shear angular power spectrum in an LSST-like survey.
We propose a mitigation method to decrease the bias in the determination of cosmological parameters due to residual errors in the 
baryonic modelling that uses the halo model-based {\tt HMCode} \citep{Mead:2015yca}\footnote{During the completion of this work a new version of HMCode was released \citep{Mead_2021}. 
This new version includes gas expulsion by AGN feedback and encapsulates star formation. Different, more physical parameters are introduced. 
The study of the consequences of the new code to our analysis is beyond the scope of the present work. }.
This method is based on an extended covariance matrix that incorporates baryonic uncertainties informed by hydrodynamical simulations.

The extended covariance matrix is constructed using  the residual errors in the best-fit modelling of 13 hydrodynamical simulations using {\tt HMCode}.
We interpret these residual errors as a random variable and integrate over them to obtain the extended covariance matrix. Nevertheless, there is some freedom in this interpretation and
therefore we studied three different possibilities for what is called the envelope of the residual errors: the Mirror, Variance and 2Mirror envelopes.  
 Although, we do not provide the ultimate prescription for how to robustly estimate the envelope of the theoretical error, the results from Fig. \ref{fig:1dA} and Table \ref{table:1d constraints} show that the use of these extended covariance matrices can lead to a significant reduction in the bias of the estimated cosmological parameters at the cost of a small increase in the uncertainties in the
parameters. The proposed choices about baryonic errors are still very dependent on the set of simulations. 
  How to optimize theoretical error without unnecessarily down-weighting any data points and in an as model-independent way as possible remains one of the main open questions that deserves more investigation in the future. Finding the answer is the key for having a truly robust analysis with reliable error bars on cosmological parameters.

It must be also emphasized that the presented analysis is a first exploratory investigation, as many simplifications were assumed such as a reduced space of cosmological parameters, neglecting non-gaussian contributions to the fiducial covariance matrix and not including other sources of systematic effects, e.g., intrinsic alignments.

Our results are encouraging since for the scenarios studied in this paper, the reduction in the residual bias consistently compensates for the increase in the statistical error. Furthermore, the proposed method is easy to implement and computationally inexpensive, providing an interesting alternative to the more conservative scale-cut methods. 
We conclude that mitigation method for baryonic uncertainties described here is a promising and viable option for analyzing data with the quality level expected at the future surveys like LSST and it deserves further investigation.

\section{acknowledgements}
This research was partially supported by the Laborat\'orio Interinstitucional de e-Astronomia (LIneA), the Brazilian funding agencies  CNPq and CAPES,
the  Instituto Nacional de Ci\^{e}ncia e Tecnologia (INCT) e-Universe (CNPq grant 465376/2014-2), and FAPESP (grant 2016/01343-7). XF is supported by the NASA ROSES ATP 16-ATP16-0084 grant, and EK is supported by the Department of Energy grant DE-SC0020247 and the David \& Lucile Packard Foundation.
The authors acknowledge the use of computational resources from LIneA, the Center for Scientific Computing (NCC/GridUNESP) of the Sao Paulo State University (UNESP), from the National Laboratory for Scientific Computing (LNCC/MCTI, Brazil), where the SDumont supercomputer ({\tt sdumont.lncc.br}) was used, and from the High Performance Computing (HPC) resources supported by the University of Arizona TRIF, UITS, and RDI and maintained by the UA Research Technologies department.

\bibliographystyle{mnras}
\bibliography{literature}

\end{document}